\documentclass[10pt]{article}
\usepackage{latexsym}
\usepackage{amssymb}
\usepackage{amsfonts}
\usepackage{amsmath}
\usepackage{theorem}

\theorembodyfont{\upshape}

\newtheorem{theorem}{Theorem}
\newtheorem{lemma}{Lemma}

\newtheorem{definition}{Definition}

\newtheorem{proposition}{Proposition}

\begin{document}

\title{Ultrametric Broken Replica Symmetry RaMOSt}
\author{Luca De Sanctis
\footnote{ICTP, Strada Costiera 11, 34014 Trieste, Italy,
{\tt<lde\_sanc@ictp.it>}}}

\maketitle

\begin{abstract}
We propose an ultrametric breaking of replica symmetry
for diluted spin glasses in the framework of Random Multi-Overlap Structures
(RaMOSt). 
Our approach permits to bound the free energy
through a trial function that depends
on a set of numbers over which one has to take the infimum.   
Such trial function is a first (ultrametric and factorized) example of a bound
in the intersection of the probability spaces of the iterative and the 
RaMOSt theories, and it shows that a ``direct dilution" of the Parisi 
Ansatz is not always exact.
\end{abstract}

\noindent{\em Key words and phrases:} diluted 
spin glasses, replica symmetry breaking, ultrametric overlap structures.

\section{Introdution}

In the case of non-diluted spin glasses, M. Aizenman R. Sims and 
S. L. Starr$^{\cite{ass}}$
introduced the idea of Random Overlap Structure (ROSt) 
to express in a very elegant manner
the free energy of the model as an infimum over a rich probability space,
to exhibit an optimal structure (the so-called Boltzmann one), to write down a
general trial function through which one can 
formulate various ansatz's for the free energy of the model.
It was also described how to formulate in particular 
the Parisi ansatz within this formalism. 

In the context of diluted spin glasses, M. Mezard and G. Parisi$^{\cite{parisi2}}$
showed how to implement the Replica Symmetry Breaking theory by translating it
into the iterative approach. 
The result is that the Broken Replica Symmetry trial function
depends on a nested chain (of Parisi type) of probabilty distributions and the order
parameter is always a function.
The rigorous proof that the Replica Symmetry Breaking in the sense of ref. \cite{parisi2}
yields bounds for the free energy has been given by 
S. Franz and M. Leone$^{\cite{franz}}$, and 
D. Panchenko and M. Talagrand$^{\cite{t1}}$.

In ref. \cite{lds1} we extended the concept of ROSt to the one 
of Random Multi-Overlap Structure (RaMOSt),
to deal with diluted spin glasses. Like in the non-diluted case, we could
express the free energy of the model by means of the Extended Variational Principle,
exhibit the optimal Boltzmann RaMOSt, 
write down the generic trial function, find a factorization property of the optimal
structures.
Here we extend in a natural way the Parisi ROSt to an ultrametric
RaMOSt. This is interesting to do to check whether
a ``minimal dilution" of the Parisi theory can be valid. It turns out
that such a minimal dilution leads to a trial function that 
is properly factorized and exact in some
regions, not exact in
some others (while in some of these regions the iterative
method yields exact results). This means that dilute models
are deeply different from their infinite connectivity limit, 
but it is not clear how.

The physics implied by the RaMOSt theory is  
different (simpler) from the one implied by the iterative method,
as it is determined explicitly and entirely by the multi-overlaps
and the trial functions depend only on a set of numbers
(fixed trial multi-overlaps) and not functions (like the distribution of the primary
fields of the iterative approach).
The ultrametric trial function we suggest, is the restriction of the bound
proved in refs. \cite{franz, t1} to the simpler framework of RaMOSt, but
our approach has nothing to share with the
iterative method. Nonetheless, our approach allows us to
recover the formulas and bounds from the iterative approach in a different
and very simple way, provided one leaves certain 
random variables to be generic instead of
making the restriction that we will use to 
impose ultrametricity in a multi-overlap structure (this connection between
the two methods will be clearer later on).
One way to look at the two approaches is the following.
One has to introduce some variables to infimize over on which
the trial free energy depends on. The iterative method considers the
cavity fields generic random variables and let their distribution vary.
The RaMOSt theory let the chosen trial multi-overlaps and 
their (probability) weights vary. In the iterative approach the weights are
of a given form that cannot change. In the RaMOSt approach the cavity
fields obey a certain constraint. We will 
point out some advantages of the RaMOSt theory, some weak
points, some open problems.

We start our treatment by illustrating the Replica Symmetric bound (in section \ref{rs}), 
in a very simple way, close to the strategy typically used for non-diluted systems.
In this simpler setting we can easily introduce all the ideas we need in the general scheme that we report in detail in section \ref{rsb}. 
The physical ideas at the basis of our bound 
are suggested by a particular interpretation 
of the Parisi theory for the (non-dilute) SK model which we 
describe in the last Appendices.
 

\section{Model, Notations, Definitions}

Notations: \\
$\alpha, \beta$ are 
non-negative real numbers
(degree of connectivity and inverse temperature
respectively);\\
$P_\zeta$ is a Poisson random variable of mean $\zeta$; \\
$\{i_\nu\}, \{j_\nu\}$ are independent identically 
distributed random variables, 
uniformly distributed over points $\{1,\ldots, N\}$;\\
$\{J_\nu\}, J$ are
independent identically distributed 
random variables, with symmetric distribution;\\
$\{\tilde{J}_\nu\}$ are
independent identically distributed 
random variables, with symmetric distribution
(different from that of $J$);\\
$\mathcal{J}$ is the set of all the quenched 
random variables  above;\\
$\sigma: i \rightarrow \sigma_{i}$ is a spin 
configuration;\\
$\pi_{\zeta}(\cdot)$ is the Poisson measure
of mean $\zeta$;\\ 
$\mathbb{E}$ is an average over all (or some of) the 
quenched variables;\\
$\omega_\mathcal{J}$ is the Bolztmann-Gibbs 
average explicitly written below;\\
$\Omega_N$ is a product of the needed 
number of independent identical copies (replicas) 
of $\omega_\mathcal{J}$;\\
$\langle\cdot\rangle$ will indicate the composition 
of an $\mathbb{E}$-type average over some 
quenched variables and some sort of 
Boltzmann-Gibbs average over the spin variables, 
that will be clear from the context.\\
We will often drop the dependance on 
some variables or indices or slightly
change notations to lighten 
the expressions, 
when there is no ambiguity.\\
We will consider only the case of zero external field, and hence
the Hamiltonian of the system of $N$ sites
is, by definition
\begin{equation*}
\label{ham}
H^{VB}_N(\sigma, \alpha; \mathcal{J})=
-\sum_{\nu=1}^{P_{\alpha N}} J_\nu \sigma_{i_\nu}\sigma_{j_\nu}
\end{equation*}
We follow the usual basic definitions and notations 
of thermodynamics for the partition function and 
the free energy per site 
\begin{eqnarray*}
\label{z}
&&Z_N(H^{VB}_{N}; \beta, \alpha; \mathcal{J})=\sum_{\{\sigma\}}
\exp(-\beta H^{VB}_N(\sigma, \alpha; \mathcal{J})),\\
\label{f}
&&-\beta f_N(\beta, \alpha)=\frac1N \mathbb{E}
\ln Z_N(\beta, \alpha; \mathcal{J})
\end{eqnarray*}
and $f=\lim_N f_N$.\\
The Boltzmann-Gibbs average of an observable $\mathcal{O}$ is
\begin{equation*}
\omega_{\mathcal{J}}(\mathcal{O})=
Z_N(\beta, \alpha; \mathcal{J})^{-1}
\sum_{\{\sigma\}}\mathcal{O}(\sigma)\exp(-\beta
H^{VB}_N(\sigma,\alpha;\mathcal{J}))
\end{equation*}
The multi-overlaps are defined (using replicas) by
\begin{equation*}
\label{overlap}
q_{n}=\frac{1}{N}\sum_{i=1}^N\sigma_i^{(1)}\cdots\sigma_i^{(n)}  
=q_{1\cdots n}
\end{equation*}
 \begin{definition} 
 A {\bf Random Multi-Overlap Structure} 
$\mathcal{R}$ is a triple 
$(\Sigma, \{\tilde{q}_{2n}\}, \xi)$ where  
\begin{itemize}
  \item $\Sigma$ is a discrete space;
  \item $\xi: \Sigma\rightarrow\mathbb{R}_+$ 
  is a system of random weights; 
  \item $\tilde{q}_{2n}:\Sigma^{2n}\rightarrow[0, 1] , n\in\mathbb{N} , |\tilde{q}|\leq 1$
  is a  positive definite \emph{Multi-Overlap Kernel} 
  (equal to 1 only on the diagonal of
  $\Sigma^{2n}$).
\end{itemize}
\end{definition}
Notice that the RaMOSt just defined is the {\sl minimal} extension of 
the concept of ROSt
to a case where all even multi-overlaps must be considered. This is quite the case
when dealing with diluted spin glasses, as a consequence of the fact
that here the distribution of the coupling is generic and hence determined by all
its moments, while in the non-diluted case the couplings are centered
Gaussians and thus determined by the second moment only. That is why all the calculations
that in the SK case end up in a single term with the 2-overlaps are replaced
here by series with all (even) multi-overlaps. So the SK case can be seen as the one where
the series stops at the first term (equivalently, as the infinite connectivity limit) and
hence we have a recipe to translate from infinite to finite connectivity and vice versa
(diluting), modulo a proper temperature rescaling.


\section{Previous results}\label{pr}

Notice that$^{\cite{lds1}}$, with $H_{N}^{VB}=H$
\begin{equation}
\label{ }
\frac{d}{d\alpha}\frac{1}{N}\mathbb{E}\ln \sum_{\gamma}\xi_{\gamma}
\exp(-\beta H)
= \sum_{n>0}\frac{1}{2n}\mathbb{E}\tanh^{2n}(\beta J)
(1-\langle q^{2}_{2n}\rangle)\ .
\end{equation}
Consider two random variables 
$\tilde{H}_{.}(\gamma, \alpha; \tilde{J})$ and $\hat{H}(\gamma, \alpha; \hat{J})$ 
such that
\begin{eqnarray}
\frac{d}{d\alpha}\mathbb{E}\ln \sum_{\gamma}\xi_{\gamma}
\exp(-\beta\tilde{H}_{.})
& = & 2\sum_{n>0}\frac{1}{2n}\mathbb{E}\tanh^{2n}(\beta J)
(1-\langle \tilde{q}_{2n}\rangle)\label{eta} \\
\frac{d}{d\alpha}\frac 1N\mathbb{E}\ln \sum_{\gamma}\xi_{\gamma}
\exp(-\beta \hat{H})
&=& \sum_{n>0}\frac{1}{2n}\mathbb{E}\tanh^{2n}(\beta J)
(1-\langle\tilde{q}^{2}_{2n}\rangle)\label{kappa}
\end{eqnarray}
and the trial function
\begin{equation*}
G_{N}(\mathcal{R})=\frac 1N\mathbb{E}\ln
\frac{\sum_{\sigma, \tau}\xi_{\tau}
\exp(-\beta\sum_{i=1}^{N}\tilde{H}_{i}\sigma_{i})}{\sum_{\tau}\xi_{\tau}
\exp(-\beta \hat{H})}
\end{equation*}
where $\tilde{H}_{i}$ are independent copies of $\tilde{H}_{.}$.
Then in ref. \cite{lds1} we proved the following
\begin{theorem}[Generalized Bound]
\label{b}
\begin{equation*}
\label{ }
-\beta f\leq \lim_{N\rightarrow\infty}
\inf_{\mathcal{R}} G_N(\mathcal{R})\ ;
\end{equation*}
\end{theorem}
\begin{theorem}[Extended Variational Principle]
\begin{equation*}
\label{evp}
-\beta f=\lim_{N\rightarrow\infty}
\inf_{\mathcal{R}}G_{N}(\mathcal{R})\ ;
\end{equation*}
\end{theorem}
\begin{theorem}[Factorization of optimal RaMOSt's]
\label{lisboa}
In the whole region where the parameters are uniquely
defined, the following Ces{\`a}ro  limit 
is linear in $N$ and $\bar{\alpha}$
\begin{equation*}\label{limrost}
\mathbf{C}\lim_{M}\mathbb{E}\ln\Omega_M
\{\sum_{\sigma}\exp[-\beta(\tilde{H}(\alpha)+\hat{H}(\bar{\alpha}/N))]\}
=N(-\beta f +\alpha A)+\bar{\alpha}A\ ,
\end{equation*}
where
\begin{equation*}
\label{ }
A=\sum_{n=1}^{\infty}\frac{1}{2n}
\mathbb{E}\tanh^{2n}(\beta J)(1-\langle q_{2n}^{2}\rangle)\ ,\ 
\tilde{H}=\sum_{i=1}^{N}\tilde{H}_{i}\sigma_{i}
\end{equation*}
\end{theorem}
We will see in the next sections what are the conditions
on $\tilde{H}_{.}$ and $\hat{H}$ in order to obey ({\ref{eta})-(\ref{kappa})
when they have a form similar to the Viana-Bray Hamiltonian. Concretely,
it is like in the non-diluted case, where the two variables are always the same,
just realized in different spaces. In other words, what really changes is
$\Sigma$, and $\tilde{H}_{.}$ and $\hat{H}$ assume different representations
accordingly.

We want now to construct a trial function with some features. We want it to 
satisfy the invariance property of the optimal structures, we want it 
to be some dilution of the Parisi trial function for the SK model and
to implement ultrametric breaking of replica symmetry, we want it
to depend on the distribution of the original couplings only, as physically
we do not expect other probability distributions to play any role,
we want it to connect the iterative method with the RaMOSt theory, by
being a restriction of the general
trial function of the iterative approach.


\section{The Replica Symmetric RaMOSt}\label{rs}

In this section we find the Replica Symmetric trial function within the
RaMOSt approach, with no external field.

The choice of the probability space of the Replica Symmetric
RaMOSt  is trivial, as we do not really need it,
just like in the non-diluted case. Still, it will serve as a guide to
the next section.

Here is the interpolating Hamiltonian
\begin{equation*}
\label{ }
H(t)=
-\sum_{\nu=1}^{P_{\alpha tN}}J_{\nu}\sigma_{i_{\nu}}\sigma_{j_{\nu}}
-\sum_{\nu=1}^{P_{2(1-t)\alpha N}}\tilde{J}_{\nu}\sigma_{i_{\nu}}
-\sum_{\nu=1}^{P_{\alpha t N}}\hat{J}_{\nu}
\end{equation*}
where $\tilde{J}_{.}$ and $\hat{J}_{.}$ are 
symmetric random variables which might have
different distribution (and different from the one of $J_{.}$).
The partition function $Z(t)$ associated to this 
Hamiltonian is defined in the usual way
and the usual derivative yields the following 
standard calculation (see e.g. \cite{lds1})
\begin{eqnarray*}
\frac{d}{dt}\frac{1}{N}\mathbb{E}\ln Z_{N}(t)&=&\alpha\mathbb{E}\ln\cosh(\beta J)
-2\alpha\mathbb{E}\ln\cosh(\beta \tilde{J})+\alpha\mathbb{E}\ln\cosh(\beta \hat{J})\\
&{}&-\alpha\sum_{n>0}\frac{1}{2n}\langle\tanh^{2n}(\beta J)
 q^{2}_{2n}-2q_{2n}\tanh^{2n}(\beta \tilde{J})\\
 &{}&\hspace{5cm}+\tanh^{2n}(\beta\hat{J})\rangle_{t}\ .
\end{eqnarray*}
where the term in $\hat{J}$ is clearly vanishing (if $\hat{J}$ is symmetric)
but we put it there
because we wanted to add and subtract a certain quantity written in two different
ways trying to ``compose a square''. Expressing the exponential 
of the part in $\hat{J}$ in terms
of hyperbolic cosine and tangent (as opposed to just cancel it trivially 
with the logarithm) complicates things but yields 
the right expressions for the quantity to be added and subtracted. 
The contribution at $t=0$ to the $t$-dependent free energy
is computed in Appendix \ref{zero}. From the expression above it is 
clear that the order parameter has to be determined by 
$\tanh^{2n}(\beta \tilde{J})/\tanh^{2n}(\beta J)$. It is therefore convenient to 
give such fractions a name by defining the so-called primary field $g$
so that
$$
\tanh(\beta \tilde{J})=\tanh(\beta J)\tanh(\beta g)\ .
$$
One can readily check that using this definition the next steps 
lead to the usual Replica Symmetric 
trial function which also gives
the correct critical point if expanded in power
series (at the fourth order).
We want instead to perform a specific choice in order to
include the Replica Symmetric trial function
within the framework of RaMOSt's. Namely
let us choose $\tilde{J}$ and $\hat{J}$ such that
\begin{equation}
\label{choice}
\tanh(\beta \tilde{J})=\tanh(\beta J)\tilde{\omega}_{\tilde{\alpha}}(\rho_{k_{\nu}})\ ,\
\tanh(\beta \hat{J})=\tanh(\beta J)\tilde{\omega}_{\tilde{\alpha}}(\rho_{k_{\nu}})
\tilde{\omega}_{\tilde{\alpha}}(\rho_{l_{\nu}})
\end{equation}
where $\tilde{\omega}_{\tilde{\alpha}}(\rho_{k_{\nu}})$ is the infinite volume limit
of the Boltzmann-Gibbs 
average of a random spin from an auxiliary system with a
Viana-Bray one-body interaction Hamiltonian
at connectivity $\tilde{\alpha}$. 
This new system has 
spins denoted by $\rho_{k}$, multi-overlaps denoted by $\tilde{q}_{2n}$, 
same couplings (independent copies) as the ones of the original system. 
Notice that 
given any
trial multi-overlap 
there exists $\tilde{\alpha}$ such that the averaged multi-overlap take that value
(see Appendix \ref{zero}).
From our choice it is clear that a single $\tilde{\alpha}$ generates a whole
sequence $\{\tilde{q}_{2n}(\tilde{\alpha})\}$ of trial multi-overlaps. Our
approach is based on the assumption that we can limit our trial functions to
such sequences. 
Notice that the distribution of of $\tilde{J}$ is 
completely determined by the one of $J$
only, as no other quenched couplings arise.\\
Now we can use the identities
\begin{equation}
\label{log}
\ln\cosh(\cdot)=\sum_{n=1}^{\infty}\frac{1}{2n}\tanh^{2n}(\cdot)\ ,
\mathbb{E}\ \tilde{\omega}_{\tilde{\alpha}}^{2n}
(\rho_{k_{.}})=\langle \tilde{q}_{2n}\rangle_{\tilde{\alpha}}=\tilde{q}_{2n}(\tilde{\alpha})
\end{equation}
to verify that the terms in $\hat{J}$ actually mutually cancel out
and also to get
\begin{equation*}
\frac{d}{dt}\frac{1}{N}\mathbb{E}\ln Z_{N}(t)=\alpha
\sum_{n>0}\frac{1}{2n}\mathbb{E}\tanh^{2n}(\beta J)
\langle (1-\tilde{q}_{2n}(\tilde{\alpha}))^{2}
-(q_{2n}-\tilde{q}_{2n}(\tilde{\alpha}))^2\rangle_{t}
\end{equation*}
where the $t$-dependent expectation has definite sign, hence we obtain
the Replica Symmetric bound and trial function from the 
fundamental theorem of calculus
and Lemma \ref{lemma} 
\begin{eqnarray*}
&F_{RS}(\beta, \alpha; \{\tilde{q}_{2n}(\tilde{\alpha})\})&=\ln 2 +
\mathbb{E}\ln\cosh(\beta\sum_{\nu=1}^{P_{2\alpha}}\tilde{J}_{\nu}(\tilde{\alpha}))\\
&{}&\hspace{0.5cm}+\alpha\sum_{n=1}^{\infty}\frac{1}{2n}\mathbb{E}
\tanh^{2n}(\beta J)(1-\tilde{q}_{2n}(\tilde{\alpha}))^{2}
\end{eqnarray*}
which is the restriction of the usual Replica Symmetric trial function
to the choice (\ref{choice}) and gives the correct annealed
solution for $\tilde{\alpha}=0$, since $\tilde{q}_{2n}(0)=0$ and
$\tilde{J}(0)=0$. In other words, what we did is to conjecture that we can limit
the trial function to those primary fields $g$ with moments (of $\tanh(\beta g)$)
satisfying a certain constraint, namely the one given by the multi-overlaps. The
reason relies on the Extended Variational Principle of ref. 
\cite{lds1} and the results of the next
section. After all, on a physical basis we do not 
expect the occurrence of other probability
distributions different from and totally independent of 
that of the original couplings.

We want now to get the whole trial function in the value at zero
of the ``interpolating pressure'' and we want to be left with a definite sign 
derivative yielding an immediate bound. This becomes essential in 
the Replica Symmetry Breaking.
The interpolating Hamiltonian is  
\begin{eqnarray}\label{tham}
H(t) &=&
-\sum_{\nu=1}^{P_{\alpha tN}}J_{\nu}\sigma_{i_{\nu}}\sigma_{j_{\nu}}
-\sum_{\nu=1}^{P_{2(1-t)\alpha N}}\left(\frac{1}{\beta}
\ln\frac{\cosh(\beta J)}{\cosh(\beta\tilde{J}_{\nu})}
+\tilde{J}_{\nu}\sigma_{i_{\nu}}\right)\nonumber \\
{}&{}&\hspace{3cm}-\sum_{\nu=1}^{P_{\alpha t N}}
\left(\frac{1}{\beta}
\ln\frac{\cosh(\beta J)}{\cosh(\beta\hat{J}_{\nu})}
+\hat{J}_{\nu}\right)
\end{eqnarray}
and the generalized trial function is
\begin{equation}
\label{trial}
G_{N}=\frac{1}{N}\mathbb{E}\ln\frac{\sum_{\sigma}
\exp(-\beta\tilde{H}(\sigma))}{\exp(-\beta\hat{H})}=
R(0)
\end{equation}
where
\begin{equation}\label{rschoice}
\tilde{H}(\sigma)= H(0) \ , \ \hat{H}=H(1)-H^{VB}_{N}(\sigma)\ , \ 
R(t)=\frac{1}{N}\mathbb{E}\ln\frac{Z(t)}{\exp(-\beta\hat{H})}
\end{equation}
and $Z(t)=Z(H(t))$ is defined in the usual way.\\
Now from Lemma \ref{lemma} in Appendix \ref{zero} we get
\begin{eqnarray*}
G_{N}=G_{RS}(\tilde{\alpha})&=&\ln 2 + \mathbb{E}\ln\cosh(\beta\sum_{\nu=1}^{P_{2\alpha}}\tilde{J}_{\nu})\\
{}&{}& + \alpha\mathbb{E}\ln\cosh(\beta J)-2\alpha\mathbb{E}\ln\cosh(\beta \tilde{J})
+\alpha\mathbb{E}\ln\cosh(\beta \hat{J})
\end{eqnarray*}
and
\begin{multline*}
\frac{d}{dt}\frac{1}{N}\mathbb{E}\ln\frac{Z(t)}{\exp(-\beta\hat{H})}=\\
-\alpha\sum_{n>0}\frac{1}{2n}\langle\tanh^{2n}(\beta J)
 q^{2}_{2n}-2q_{2n}\tanh^{2n}(\beta \tilde{J})
 +\tanh^{2n}(\beta\hat{J})\rangle_{t}
\end{multline*}
becomes
 \begin{equation*}
\frac{d}{dt}\frac{1}{N}\mathbb{E}\ln\frac{Z(t)}{\exp(-\beta\hat{H})}=-\alpha
\sum_{n>0}\frac{1}{2n}\mathbb{E}\tanh^{2n}(\beta J)
\langle(q_{2n}-
\tilde{q}_{2n}(\tilde{\alpha}))^2\rangle_{t}
\end{equation*}
with the usual choices for $\tilde{J}_{.}$ and $\hat{J}_{.}$.\\
Since
\begin{equation*}
\frac{1}{N}\mathbb{E}\ln\frac{Z(1)}{\exp(-\beta\hat{H})}=-\beta f^{VB}_{N}
\end{equation*}
the fundamental theorem of calculus and the definite sign of the derivative above
provide again the Replica Symmetric trial function and bound that we summarize
in the just proved 
\begin{theorem}
With the choice defined by $(\ref{trial})$, $(\ref{rschoice})$, and $(\ref{tham})$,
$\tilde{H}_.$ and $\hat{H}$ satisfy $(\ref{eta})-(\ref{kappa})$, and 
\begin{equation*}
\label{ }
-\beta f(\beta, \alpha) \leq G_{RS} (\tilde{\alpha})
= F_{RS}(\tilde{\alpha})\ \ \ \forall\ \tilde{\alpha}\
\in [0, \infty]\ .
\end{equation*}
\end{theorem}

Notice that, if $\beta^{\prime 2}=\alpha\mathbb{E}\tanh^2(\beta J)$ 
is fixed, then
\begin{equation*}
\label{ }
\lim_{\alpha \to \infty}F^{VB}_{RS}(\beta, \alpha, \tilde{\alpha})=
F^{SK}_{RS}(\beta^{\prime})
\end{equation*}
where $F^{VB}_{RS}$ is the Replica Symmetric 
trial function for the Viana-Bray model constructed in this section, and 
$F^{SK}_{RS}$ is the Replica Symmetric trial function for
SK model.

In the non-dilute case, a complete control of the high temperature
regime can be gained by means of the quadratic replica coupling
method$^{\cite{gt2}}$, also when there is an external field. The extension of
that method to the Viana-Bray model lead the same result, only when
there is no external field$^{\cite{gt1}}$. An external field reveals an intrinsic
pathology of dilute models. But the presence of an external field seems to
be pathological for the goodness of the bounds
also in the approach we propose in this article, we will
comment on this later on. That is why we limit ourselves to the case 
of zero external field, although mathematically it would be very easy to include it
in the treatment.


\section{Replica Symmetry Breaking and Ultrametric RaMOSt}\label{rsb}

In this section we will construct a trial free energy of Parisi type
depending on ultrametric trial multi-overlaps. 
The purpose is to show that the iterative and the RaMOSt theories
can be compatible, in the sense that there are trial functions that
live both in a RaMOSt and in the probability space of the general
trial function of the iterative method. Moreover, we want to show how one 
can construct a trial function depending on ultrametric multi-overlaps, extending
the Parisi ultrametricity to the diluted case. Our main goal thus is not to 
find the exact value of the free energy, nor to get closer to it than one can get
with the iterative trial function. In fact, we will construct a trial function that turns out
to be some restriction of the iterative trial function. 
Hence the trial function in this section
cannot be closer to the true free energy than the iterative one. 

We need to generalize the ideas of the previous section,
in particular the second identity in (\ref{log}) and the preceding
discussion.
Given any partition $\{x^{a}\}_{a=0}^{K}$ of 
the interval $[0 , 1]$, there exists a sequence 
$\{\tilde{\alpha}_{a}\}_{a=0}^{K}\in[0 , \infty]$ such that
$\tilde{q}_{2n}(\tilde{\alpha}_{a})=x_{a}-x_{a-1}$. In other words,
a sequence $\{\tilde{\alpha}_{a}\}_{a=0}^{K}\in[0 , \infty]$ generates
for each $n\in \mathbb{N}$ a partition of $[0 , 1]$ considered as 
the set of trial values of $\tilde{q}_{2n}$, provided the $\tilde{\alpha}_{a}$
are not too large
\begin{equation}\label{constraint}
\sum_{a\leq K}\tilde{q}_{2n}(\tilde{\alpha}_{a})\leq 1\ .
\end{equation}
Again, we limit our trial 
multi-overlaps to belong to partitions generated in this way.
This implies that the points of the generated partitions tend to
get closer to zero as $n$ increases. This is good, since
in any probability space $\langle \tilde{q}_{2n}\rangle$ decreases
as $n$ increases and therefore the probability integral 
distribution functions tend to grow faster near zero.\\
We can then define $\tilde{W}_{\gamma}$, 
for $\gamma\in\mathbb{N}^K$, through
\begin{equation*}
\label{ }
\tilde{W}_{\gamma}(\bar{J}, k_{\nu})=
\tilde{\omega}_{\tilde{\alpha}_{1}}(\rho_{k_{\nu}})\bar{J}_{\gamma_{1}}+\cdots+
\tilde{\omega}_{\tilde{\alpha}_{K}}(\rho_{k_{\nu}})\bar{J}_{\gamma_{1}\cdots\gamma_{K}}
\end{equation*}
with $\bar{J}_{.}=\pm 1$ independent identically distributed symmetric
random variables.
\begin{definition}
$$\tilde{q}_{\gamma^{1}\cdots\gamma^{2n}}=
(\tilde{q}^{1}_{2n}-\tilde{q}^{0}_{2n})\delta_{\gamma^{1}_{1}\cdots\gamma^{2n}_{1}}
+\cdots+(\tilde{q}^{K}_{2n}-\tilde{q}^{K-1}_{2n}) \delta_{\gamma^{1}_{1}\cdots\gamma^{2n}_{1}} \cdots\delta_{\gamma^{1}_{K}\cdots\gamma^{2n}_{K}}$$ 
is the ultrametric $2n$-overlap. 
\end{definition}
Clearly this is just a kind of ultrametricity, imposed for each (even) number of 
replica {\em individually}.

The choice of $\tilde{W}_{\gamma}$ imposes an ultrametric structure since
\begin{eqnarray*}
 \mathbb{E}(\tilde{W}_{\gamma^{1}}\!\!\!\! & \cdots &\!\!\!\!\tilde{W}_{\gamma^{2n}}) = \\
 {}&=&\mathbb{E}\tilde{\omega}^{2n}_{\tilde{\alpha}_{1}}(\rho_{.})
 \delta_{\gamma^{1}_{1}\cdots\gamma^{2n}_{1}}+
 \cdots+\mathbb{E}\tilde{\omega}^{2n}_{\tilde{\alpha}_{K}}(\rho_{.})
 \delta_{\gamma^{1}_{1}\cdots\gamma^{2n}_{1}}
 \cdots\delta_{\gamma^{1}_{K}\cdots\gamma^{2n}_{K}} \\
{} & = & \tilde{q}_{2n}(\tilde{\alpha}_{1})\delta_{\gamma^{1}_{1}\cdots\gamma^{2n}_{1}} 
+\cdots+\tilde{q}_{2n}(\tilde{\alpha}_{K}) \delta_{\gamma^{1}_{1}\cdots\gamma^{2n}_{1}}
 \cdots\delta_{\gamma^{1}_{K}\cdots\gamma^{2n}_{K}}\\
 {} & = & (\tilde{q}^{1}_{2n}-\tilde{q}^{0}_{2n})\delta_{\gamma^{1}_{1}\cdots\gamma^{2n}_{1}}
+\cdots+(\tilde{q}^{K}_{2n}-\tilde{q}^{K-1}_{2n}) \delta_{\gamma^{1}_{1}\cdots\gamma^{2n}_{1}} \cdots\delta_{\gamma^{1}_{K}\cdots\gamma^{2n}_{K}} \\
{} & \equiv & \tilde{q}_{2n}=\tilde{q}_{\gamma^{1}\cdots\gamma^{2n}}\ .
\end{eqnarray*}
If we also define
\begin{equation*}
\label{ }
\hat{W}_{\gamma}=\tilde{W}_{\gamma}(\bar{J}, k_{\nu})
\tilde{W}_{\gamma}(\bar{J}^{\prime}, l_{\nu})
\end{equation*}
where $\bar{J}^{\prime}$ denotes independent copies of $\bar{J}$, 
we have
\begin{multline*}
 \mathbb{E}(\hat{W}_{\gamma^{1}}\cdots\hat{W}_{\gamma^{2n}})
 = \tilde{q}^{2}_{2n}=\tilde{q}^{2}_{\gamma^{1}\cdots\gamma^{2n}}=\\
[(\tilde{q}^{1}_{2n})^{2}-(\tilde{q}^{0}_{2n})^{2}]\delta_{\gamma^{1}_{1}\cdots\gamma^{2n}_{1}}
+\cdots+[(\tilde{q}^{K}_{2n})^{2}-(\tilde{q}^{K-1}_{2n})^{2}] 
\delta_{\gamma^{1}_{1}\cdots\gamma^{2n}_{1}} 
\cdots\delta_{\gamma^{1}_{K}\cdots\gamma^{2n}_{K}}  
\end{multline*}
where the last expected equality can be easily verified by direct calculation.
We clearly have in mind the case $\tilde{q}^{0}_{2n}=0, \tilde{q}^{K}_{2n}=1$
(which implies the equal sign holds in (\ref{constraint})).

Now given a set of weights $\xi_{\gamma}, \gamma\in\mathbb{N}^K$,
we can state the next
\begin{proposition}
There exist $\tilde{H}, \hat{H}$ satisfying (\ref{eta})-(\ref{kappa}) with 
$\tilde{q}$ ultrametric.
\end{proposition}
Before proving this proposition, let us 
consider the usual set of weights 
$\xi_{\gamma}(m_{1}, \ldots , m_{K}), \gamma=(\gamma_{1}, \ldots , \gamma_{K})$ 
associated to the Random Probability Cascade
of Poisson-Dirichlet Processes through which one can express formulas
of Parisi type (see e.g. ref. \cite{t1}).
Then take the trial function
\begin{equation*}
G_{N}=\frac{1}{N}\mathbb{E}\ln\sum_{\gamma, \sigma}
\xi_{\gamma}\exp(-\beta \tilde{H}_{\gamma})-
\frac{1}{N}\mathbb{E}\ln\sum_{\gamma}
\xi_{\gamma}\exp(-\beta\hat{H}_{\gamma})\ .
\end{equation*} 
Notice that the trial function above is the usual 
difference between the ``cavity" term 
and the ``internal" term.
Denoting by $X$ the map 
$$
X:\tilde{\alpha}_{a}\rightarrow m_{a}
$$
satisfying (\ref{constraint})
we can consider the trial function as a function $G(X)$ of $X$.
We will prove the proposition above together with the next
\begin{theorem} The ultrametric trial function $G(X)$ satisfies the bound
\begin{equation*}
-\beta f(\beta, \alpha)\leq\inf_{X}G(X)
\end{equation*}
as in Theorem \ref{b}, it enjoys the factorization
property
as in Theorem \ref{limrost} (in the sense that $\tilde{H}$ and $\hat{H}$ 
are independent, and each spin yields the same independent
contribution) and it reduces to the Parisi trial function
for the SK model in the infinite connectivity limit.
\end{theorem}
{\bf Proof}.
Consider the interpolating Hamiltonian 
\begin{equation*}
H_{\gamma}(t)=H^{VB}(t)+\tilde{H}_{\gamma}(1-t)+\hat{H}_{\gamma}(t)
\end{equation*}
where
\begin{eqnarray*}
\tilde{H}_{\gamma} &=&
-\sum_{\nu=1}^{P_{2\alpha N}}\left(\frac1\beta
\ln\frac{\cosh(\beta J)}{\cosh(\beta 
\tilde{J}^{\gamma}_{\nu})}
+\tilde{J}^{\gamma}_{\nu}\sigma_{i_{\nu}}\right)\\
\hat{H}_{\gamma} &=&
-\sum_{\nu=1}^{P_{\alpha N}}\left(\frac1\beta
\ln\frac{\cosh(\beta J)}{\cosh(\beta 
\hat{J}^{\gamma}_{\nu})}
+\hat{J}^{\gamma}_{\nu}\right)
\end{eqnarray*}
and $t$ is understood to multiply the connectivity $\alpha$.
Consider
\begin{equation*}
R(t)=\frac{1}{N}\mathbb{E}\ln\frac{\sum_{\gamma, \sigma}
\xi_{\gamma}\exp(-\beta H_{\gamma}(t))}{\sum_{\gamma}
\xi_{\gamma}\exp(-\beta\hat{H}_{\gamma})}\ , \ G_{N}=R(0)
\end{equation*}
This time let us chose $\tilde{J}_{\gamma}, \hat{J}_{\gamma}$ of the form
\begin{equation*}
\label{ }
\tanh(\beta\tilde{J}_{\gamma})=\tanh(\beta J)\tilde{W}_{\gamma}\ ,\ 
\tanh(\beta\hat{J}_{\gamma})=\tanh(\beta J)\hat{W}_{\gamma}
\end{equation*}
and compute the usual $t$-derivative
\begin{eqnarray*}
\frac{d}{dt}R(t) & = & \alpha\mathbb{E}\sum_{n>0}\frac{1}{2n}
\mathbb{E}\tanh^{2n}{(\beta J)}\mathbb{E}[
\Omega^{2n}_{t}(\sigma_{i_{\nu}}\sigma_{j_{\nu}}) \\
{} & {} & -2\Omega^{2n}_{t}(\tilde{W}_{\gamma}\sigma_{i_{\nu}})
+\Omega^{2n}_{t}(\hat{W}_{\gamma})]
\end{eqnarray*}
where the $\Omega_{t}$ is the generalized Boltzmann-Gibbs average with the 
weights $\xi_{.}$ and the Hamiltonian $H(t)$.\\
It is obvious that
\begin{equation*}
\label{ }
\mathbb{E}\Omega^{2n}_{t}(\tilde{W}_{\gamma}\sigma_{i_{\nu}})=
\langle \tilde{q}_{2n} q_{2n}\rangle_{t}\ ,\ 
\mathbb{E}\Omega^{2n}_{t}(\hat{W}_{\gamma})=
\langle \tilde{q}^2_{2n}\rangle_{t}\ .
\end{equation*}
This proves the proposition.

The RaMOSt is thus equipped with all the ingredients we need and
we finally obtain
\begin{equation}
\label{derivative}
\frac{d}{dt}R(t)=
-\alpha\sum_{n>0}\frac{1}{2n}\mathbb{E}\tanh^{2n}{(\beta J)}
\langle (\tilde{q}_{2n}- q_{2n})^{2}\rangle_{t}
\end{equation}
which is exactly the same expression 
as in equation (5) of ref. \cite{lds1}, except 
here the trial multi-overlaps are not the Boltzmann ones, but rather
some ultrametric ones, in the strictest analogy with the Parisi ROSt 
for SK. From (\ref{derivative}) we clearly get the ultrametric bound
that we wanted to prove.

Notice that $G(X)$ does not depend on $N$, thanks to 
the same calculations that led to Lemma
\ref{lemma} in Appendix \ref{zero}. Moreover
$\tilde{W}$ and $\hat{W}$ are chosen to be
independent, therefore the factorization property of the 
optimal RaMOSt's illustrated
in ref. \cite{lds1} holds:
$$
\mathbb{E}\ln \Omega_\xi[c_1\cdots c_N\exp(-\beta \hat{H}(\bar{\alpha}/N))]
=NB+\bar{\alpha}A
$$
for some $B$, and we used the notation (\cite{lds1})
$$
c_1\cdots c_N=\sum_\sigma\exp(-\beta \tilde{H})\ .
$$
Finally, a simple interpolation between the Parisi trial free energy
(\ref{parisirsb}), at the $K$-th level of replica symmetry breaking,
and $G(X)$, at the same level of replica symmetry breaking,
shows that in the infinite connectivity limit
the two quantities coincide (modulo the proper temperature
rescaling) since the infinite connectivity kills all the multi-overlaps
but the 2-overlap, and the latter is the same ultrametric
one in the two trial functions for each model. The trial values
of the 2-overlap are the same in both trial functions. If we choose the 
trial values for the SK model, then this determines the sequence of
trial auxiliary connectivities to be used in the VB model. Vice versa,
given a sequence of auxiliary connectivities, the dilute trial function
reduces to the SK one with the same values of the trial 2-overlap.
In more general and abstract terms, the constraints (\ref{eta})-(\ref{kappa}) 
are such that a RaMOSt reduces to a ROSt with the same 
overlap kernel in the infinite connectivity limit.
$\Box$

What we did is, in other words, to ``dilute"
(\ref{physics1})-(\ref{physics2}) as
\begin{eqnarray*}
&&\mathbb{E}(\tanh(\beta \tilde{J}_{\gamma^{1}})
\cdots\tanh(\beta \tilde{J}_{\gamma^{2n}}))=
\mathbb{E}\tanh^{2n}(\beta J)\tilde{q}_{\gamma^{1}\cdots\gamma^{2n}}\ ,\\
&&\mathbb{E}(\tanh(\beta \hat{J}_{\gamma^{1}})
\cdots\tanh(\beta \hat{J}_{\gamma^{2n}}))=
\mathbb{E}\tanh^{2n}(\beta J)\tilde{q}^{2}_{\gamma^{1}\cdots\gamma^{2n}}\ .
\end{eqnarray*}

Notice that $X$ together with 
$\tilde{\alpha}_{a}\rightarrow\tilde{q}^{a}_{2n}-\tilde{q}^{a-1}_{2n}$ 
induces a map 
$$
X_{2n}(q)=m_{a}\ ,\ \tilde{q}^{a-1}_{2n}\leq q < \tilde{q}^{a}_{2n}\ .
$$

As a side remark, notice that the fundamental theorem of calculus
applied to (\ref{kappa}) implies that in any RaMOSt the 
part in $\hat{H}$ of $G_{N}$ has the usual 
integral form like in the non-diluted case
$$
\alpha\sum_{n=1}^{\infty}\frac{1}{n}\mathbb{E}
\tanh^{2n}(\beta J)\int_{0}^{1}qX_{2n}(q)dq
$$ 
where $X_{2n}$ includes
the integration in $d\alpha$ (see Appendix \ref{parisi}).
In the particular case of the Boltzmann RaMOSt, 
this is the ``internal energy term"
with the Boltzmann distribution of the multi-overlaps, since it has
the integral form above even without integrating back in $d\alpha$ (see \cite{lds1}).
In the Ultrametric RaMOSt, the corresponding distribution $X_{2n}$ 
is not the usual Parisi one that would yield
\begin{equation}
\label{internal}
\alpha\sum_{n=1}^{\infty}\frac{1}{2n}\mathbb{E}
\tanh^{2n}(\beta J)(1-\sum_{a}^{K}(m_{a+1}-m_{a})(q^{a}_{2n})^{2})
\end{equation}
while this is instead the case for the SK model. This means that
the physics of the model and the interpretation of the parameters
$\{m_{a}\}$ are still quite obscure. In order to make the internal 
energy part have the same form as in the Boltzmann RaMOSt,
one could consider for instance, among other possibilities, 
starting from (\ref{internal}), leaving $\hat{H}$ out of the 
interpolation and then try to deduce the proper choice of $\tilde{H}$.

Another remark. The trial function of the iterative method can be written using
the same weights that we used here$^{\cite{t1}}$. Then the cavity fields
have generic distributions, which are the parameters to infimize over. The
trial function we got could be obtained by
imposing a restriction to such parameters in order to make the iterative
trial function live in a RaMOSt (with ultrametric multi-overlaps). Hence in
our example we make a very special choice both for the distribution of the
cavity fields, and for the weights of the RaMOSt. This can easily mean
simplifying too much. In fact, when there is an external field and the 
connectivity is smaller than one, the Replica Symmetric trial function of the
iterative method is exact, while the Replica Symmetric 
trial function of the previous
section is not (it is not too difficult to prove it through an expansion if powers 
of $\alpha$. One could try
and break the Replica Symmetry in order to get
the exact value of the free energy, but it would be very unphysical and
it would give non-selfaveraging overlaps and hence a strict inequality in 
(\ref{derivative})). Despite its 
simplicity, the trial function of this section is enough to provide an example
of trial function with many good qualities. It is exact when there is no external field
in the high temperature regime, it has the invariance property of the optimal
structures, it exhibits ultrametricity, it depends on a few trial values
of the auxiliary connectivity through which
one can vary all the trial multi-overlaps simultaneously, it depends on a natural
(decreasing) sequence of trial multi-overlaps as physically expected, 
it connects the iterative and the RaMOSt 
approaches, it does not need to introduce new generic probability distributions
for the couplings, it is easier to compute than the iterative one, it gives
the Parisi trial function in the infinite connectivity limit, it belongs to a RaMOSt
(and for the RaMOSt's the Extended Variational Principle has been proven,
this is not the case in the iterative method).


\section{Conclusions}

The ROSt approach is physically very deep. One takes an auxiliary system
with weights and a trial overlap, then a cavity field and an internal field,
both Gaussian like the Hamiltonian. The two Gaussian variables
stay the same, what changes is the space where they are defined.
The trial overlap determines their 
covariance, which is the parameter to infimize over. A specific form of the
trial function, like the Parisi one, is associated to a specific choice of the ROSt
(space and weights). Letting the weights to be generic allows one to prove the 
extended variational principle.
In dilute spin glasses we have the analogous structure (RaMOSt) with
extended variational principle, and the 
iterative trial function, which is of the same form (with the same exponents)
as the Parisi trial function for the SK model. Now, there are some natural steps
to take. The iterative trial function {\sl must} be somehow reduced to a multi-overlap structure (because of section \ref{pr}). The trial function must be constructed
in a RaMOSt, and it is physically expected that the randomness of the
cavity and internal fields be determined by the distribution of the original couplings
(like for non-dilute spin glasses). It is interesting to find a minimal
extension of the Parisi ROSt and to see whether it gives a good trial function.
It is also interesting to see whether it is possible to construct a trial function
with ultrametric trial multi-overlaps. 
Well, a trial function fulfilling all these requirements (with the proper
invariance property)
has been found, in the previous section, and it 
is not exact in some regions. We think such result is interesting, though of negative
nature, and that the trial function is an instructive starting point for further
developments.
In any case, some considerations arise. The specific restriction of
the iterative cavity fields we performed could be wrong, or else the weights of the
RaMOSt might not be the right ones. The latter would be quite surprising
(and also imply that iterative method is wrong too). But if the restriction we exhibited
is wrong, it means that either ultrametricity does not take place in diluted spin glasses,
or at least it must be implemented in a radically new way, as the dilution of
the Parisi Ansatz is wrong. It is thus important to understand how the
physics of dilute spin glasses is different from the non-dilute ones, and how to 
restrict the iterative trial function.


\appendix

\section{The Cavity Energy}\label{zero}

Let us compute the 
one-body interaction free energy.
\begin{eqnarray*}
\frac{1}{N}\mathbb{E}\ln\sum_{\sigma}
\exp(\beta\sum_{\nu=1}^{P_{2\alpha N}}\tilde{J}_{\nu}\sigma_{i_{\nu}})&=&
\frac{1}{N}\mathbb{E}\ln\sum_{\sigma}\exp(\beta
\sum_{\nu=1}^{P_{2\alpha N}}\tilde{J}_{\nu}\sum_{i=1}^{N}\delta_{i,i_{\nu}}
\sigma_{i_{\nu}})\\
{}&=&\ln 2 + \frac{1}{N}\mathbb{E}\ln\prod_{i=1}^{N}\cosh
(\beta\sum_{\nu=1}^{P_{2\alpha N}}\tilde{J}_{\nu}\delta_{i,i_{\nu}})\\
{}&=&\ln 2 + \mathbb{E}\ln\cosh
(\beta\sum_{\nu=1}^{P_{2\alpha N}}\tilde{J}_{\nu}\delta_{1,i_{\nu}})\ .
\end{eqnarray*}
In the expression above, $k$ (out of $m$) of the $i_{\nu}$'s will be equal to $1$ 
with probability
$$
\binom mk \left(\frac{1}{N}\right)^{k}\left(\frac{N-1}{N}\right)^{m-k}
$$
and therefore
\begin{multline*}
\frac{1}{N}\mathbb{E}\ln Z_{N}^{(1)}=\ln 2 +\\
\sum_{m=0}^{\infty}\sum_{k=0}^{m}\left[
e^{-2\alpha(N-1)}e^{-2\alpha}\frac{1}{m!}(2\alpha)^{m-k}(2\alpha)^{k}N^{m}
\frac{m!}{k!(m-k)!}\frac{1}{N^{k}}
\frac{(N\!-1)^{m-k}}{N^{m-k}}
\right.\\
\left.\mathbb{E}\ln\cosh(\beta\sum_{\nu=1}^{k}\tilde{J}_{\nu})\right]\ .
\end{multline*}
Now the formula
\begin{equation*}
\label{ }
\sum_{m=0}^{\infty}\sum_{k=0}^{m}a_{m-k}b_{k}
=\left(\sum_{m=0}^{\infty}a_{m}\right)\left(\sum_{k=0}^{\infty}b_{k}\right)
\end{equation*}
applies here and yields the following 
\begin{lemma}\label{lemma}
\begin{equation*}
\label{ }
\frac{1}{N}\mathbb{E}\ln\sum_{\sigma}
\exp(\beta\sum_{\nu=1}^{P_{2\alpha N}}\tilde{J}_{\nu}\sigma_{i_{\nu}}) 
= \ln 2 +
\mathbb{E}\ln\cosh(\beta\sum_{\nu=1}^{P_{2\alpha}}\tilde{J}_{\nu})
\end{equation*}
\end{lemma}
which contains two of the terms of the Replica Symmetric trial functional.

\textbf{Remark}: it is easy to see that
\begin{equation*}
\frac{d}{d\alpha}\frac{1}{N}\mathbb{E}\ln\sum_{\sigma}
\exp(\beta\sum_{\nu=1}^{P_{2\alpha N}}\tilde{J}_{\nu}\sigma_{i_{\nu}}) 
=2\sum_{n>0}\frac{1}{2n}\mathbb{E}\tanh^{2n}{(\beta J)}
(1-\langle q_{2n}\rangle)
\end{equation*}
and since for any $m=0,1,\ldots$
\begin{equation*}
\frac{d}{d\alpha}\pi_{\alpha}(m)\rightarrow 0\ \  \Leftarrow\ \  \alpha\rightarrow \infty
\end{equation*}
where $\pi_{\alpha}(m)$ is the Poisson measure of mean $\alpha$, we deduce 
for all $n$
\begin{equation*}
\langle q_{2n}\rangle \rightarrow 1\ \  \Leftarrow\ \  \alpha\rightarrow \infty\ .
\end{equation*}


\section{Parisi theory of SK}\label{parisi}

\numberwithin{equation}{section}

Let us recall the well known SK Hamiltonian, which is defined as a centered
Gaussian with covariance given by an overlap
\begin{equation*}
\label{ }
H_{N}^{(SK)}(J)=-\frac{1}{\sqrt{N}}\sum_{i, j}^{1, N}J_{ij}\sigma_{i}\sigma_{j}
\end{equation*}

The cavity field $\tilde{H}_{i}$ acting on the spin $\sigma_{i}$ 
of the Parisi theory is given by
the following decomposition of $J$
\begin{equation*}
\label{}
-\tilde{H}_{i}=\tilde{J}_{\gamma}=
\sqrt{\tilde{q}_{1}}J^{\gamma_{1}}_{i}+\cdots+\sqrt{\tilde{q}_{K}-
\tilde{q}_{K-1}}J^{\gamma_{1}\cdots\gamma_{K}}_{i}\ ,\ i=1, ... , N
\end{equation*}
where the $\gamma$ indexes are the ones 
of the Random Probability Cascades of Poisson-Dirichlet processes,
used for the weights $\xi_{\gamma}(m_{1}, \ldots, m_{K})$ 
to express in a compact
way the nested expectations of Parisi formula.
The couplings of the cavity field are
related to the original ones by the trial overlaps
\begin{equation}
\label{physics1}
\mathbb{E}(\tilde{J}_{\gamma}\tilde{J}_{\gamma^{\prime}})=
\mathbb{E}(J^{2})\tilde{q}_{\gamma\gamma^{\prime}}\ .
\end{equation}
Each trial overlap 
$\tilde{q}_{a}$ from the assumed partition of $[0 , 1]$ can be obtained as
the overlap in an auxiliary system with a one-body interaction (for simplicity) 
with couplings $J$
modulated by a suitable strength $\sqrt{x_{a}}$, 
thanks to the monotone dependance
of the overlap on $x$, i.e. $\tilde{q}_{a}=\tilde{q}(x_{a})$. 
But we can also put
\begin{equation*}
\label{ }
\tilde{q}_{a}-\tilde{q}_{a-1}=
\tilde{q}(x_{a})-\tilde{q}(x_{a-1})=
\tilde{q}^{a}=\tilde{q}(x^{a})\ ,\ \sum_{a=1}^{K}\tilde{q}^{a}=1
\end{equation*}
and re-write the cavity field as
\begin{equation*}
\label{ }
-\tilde{H}_{i}=\sqrt{\tilde{q}(x^{1})}J^{\gamma_{1}}_{i}+\cdots+\sqrt{\tilde{q}(x^{K})}
J^{\gamma_{1}\cdots\gamma_{K}}_{i}\ .
\end{equation*}
The ultrametricity is intrinsic in the $\tilde{H}_{i}$'s, as can be easily checked by
their covariance, which is the only quantity that is related to both the overlap and
the generalized bound (see ref. \cite{ass}). 
The internal energy is therefore expressed introducing
(see ref. \cite{ass})
\begin{equation*}
\label{ }
-\hat{H}=\hat{J}_{\gamma}=
\sqrt{\tilde{q}^{2}_{1}}J^{\gamma_{1}}+\cdots+\sqrt{\tilde{q}^{2}_{K}-
\tilde{q}^{2}_{K-1}}J^{\gamma_{1}\cdots\gamma_{K}}\ ,
\end{equation*}
or equivalently
\begin{equation}
\label{physics2}
\mathbb{E}(\hat{J}_{\gamma}\hat{J}_{\gamma^{\prime}})=
\mathbb{E}(J^{2})\tilde{q}^{2}_{\gamma\gamma^{\prime}}\ ,
\end{equation}
and the trial function can be written as
\begin{equation}
\label{parisirsb}
G_P=\frac{1}{N}\mathbb{E}\ln
\frac{\sum_{\gamma, \sigma}\xi_\gamma
\exp(-\beta \sum_{i=1}^{N}\tilde{H}_{i}\sigma_{i})}
{\sum_{\gamma}\xi_\gamma\exp(-\beta \hat{H})}
=-\beta f_{K-BRS}(X)
\end{equation}
where $X$ is the Parisi order parameter.
Notice that there is a $y^{a}=y(x^{a})$ such that
\begin{equation*}
\label{ }
\tilde{q}^{2}_{a}-\tilde{q}^{2}_{a-1}=\tilde{q}^{2}(y^{a})
\end{equation*}
and that $\{y^{a}\}$ is determined by $\{x^{a}\}$ since
\begin{equation*}
\label{ }
\tilde{q}^{2}_{a}-\tilde{q}^{2}_{a-1}=(\tilde{q}_{a}-
\tilde{q}_{a-1})(\tilde{q}_{a}+\tilde{q}_{a-1})=
\tilde{q}(x^{a})(2\sum_{r=1}^{a}\tilde{q}(x^{r})+\tilde{q}(x^{a+1}))
\end{equation*}
so that the trial function $G$ can be expressed in terms of the $x^{a}$ only.

Moreover, it is easy to see that
\begin{equation*}
\label{ }
\frac{1}{N}\mathbb{E}\ln
\sum_{\gamma}\xi_\gamma\exp(-\beta \hat{H})=
\frac{\beta^{2}}{2}\int_{0}^{1}qX(q)dq=
\frac{\beta^{2}}{2}
\frac{1}{2}(1-\langle q^{2}\rangle)
\end{equation*} 
using for instance integration by parts or Fubini theorem. 
The second equality above
holds in full generality, for any average $\langle\cdot\rangle$ in some space
of a random variable $q$ between zero and one, the distribution of which can be
denoted by $X$. In particular, $X$ can be the one associated to the 
Boltzmann-Gibbs measure or the Parisi one: 
in the former case $\hat{H}$ is given in ref. \cite{ass},
the latter case has just been illustrated.


\section*{Acknoledgments}

The author thanks an anonymous referee for several very interesting comments.
The author also warmly thanks Yakov Sinai and Francesco Guerra
for support and encouragement, 
and gratefully acknowledges the hospitality of the Department of 
Physics at University of Rome ``La Sapienza'' (and in particular 
Giovanni Jona-Lasinio). 



\begin{thebibliography}{99}

   \bibitem{ass} M. Aizenman, R. Sims, S. L. Starr,
   \emph{An Extended Variational Principle for 
   the SK Spin-Glass Model}, 
   Phys. Rev. B \textbf{68}, 214403 (2003)

   \bibitem{lds1} L. De Sanctis, \emph{Random Multi-Overlap Structures and
   Cavity Fields in Diluted Spin Glasses}, J. Stat. Phys. \textbf{117}, 785-799 (2004)
   
   \bibitem{franz} S. Franz, M. Leone, \emph{Replica 
   bounds for optimization problems and 
   diluted spin  systems}, J. Stat. Phys.Ê \textbf{111}, 535 (2003)
   
   \bibitem{gt1} F. Guerra, F.L. Toninelli, \emph{The high temperature region
   of the Viana-Bray diluted spin glass model}, 
   J. Stat. Phys.Ê \textbf{115},  (2004)
   
   \bibitem{gt2} F. Guerra, F.L. Toninelli, \emph{Quadratic replica coupling
   in the Sherrington-Kirkpatrick mean field spin glass model}, 
   J. Math. Phys.Ê \textbf{43},  3704 (2002)
     
   \bibitem{parisi2} M. Mezard, G. Parisi, \emph{The
   Bethe lattice spin glass revisited}, Eur. Phys. J. B \textbf{20}, 217 (2001)

   \bibitem{t1} D. Panchenko, M. Talagrand,
   \emph{Bounds for diluted mean-field spin glass models}, 
   Prob. Theory Related Fields \textbf{130}, 319-336 (2004)
      
\end{thebibliography}
\end{document}